\documentclass[12pt]{article}
\usepackage{graphicx}
\usepackage{epsfig}
\usepackage{epstopdf}
\usepackage{amsmath}
\usepackage{amsfonts}
\usepackage{amssymb}
\usepackage{hyperref}
\newcommand{\be}{\begin{equation}}
\newcommand{\beq}{\begin{equation}}
\newcommand{\eeq}{\end{equation}}
\newcommand{\eq}{\begin{equation}}
\newcommand{\ee}{\end{equation}}
\newcommand{\beqa}{\begin{eqnarray}}
\newcommand{\eeqa}{\end{eqnarray}}
\newcommand{\bea}{\begin{eqnarray}}
\newcommand{\eea}{\end{eqnarray}}

\newcommand{\ie}{{\it i.e.,}\ }

\numberwithin{equation}{section}
\begin{document}

\thispagestyle{empty}

\begin{titlepage}

\begin{flushright}
{\small
%Preprint XXX
}
\end{flushright}
\vskip 1cm

\centerline{\Large \bf Exact Gravitational Dual of a Plasma Ball}

\vspace{1cm}

\centerline{
Roberto Emparan$^{a,b}$
and Giuseppe Milanesi$^{c}$}

\vskip .5cm
\centerline{$^a$\em Instituci\'o Catalana de Recerca i Estudis
Avan\c cats (ICREA),}
\centerline{\em Passeig Llu{\'\i}s Companys, 23, E-08010 Barcelona, Spain}
\centerline{$^b$\em Departament de F{\'\i}sica Fonamental, Universitat de
Barcelona,}
\centerline{\em Marti i Franqu\`es 1, E-08028 Barcelona, Spain}
\centerline{$^c$\em Departament ECM, Universitat de
Barcelona,}
\centerline{\em Marti i Franqu\`es 1, E-08028 Barcelona, Spain}
\bigskip
\centerline{\tt emparan@ub.edu, milanesi@ecm.ub.es}
\vskip 1cm

\vskip 1cm

\begin{abstract}

\noindent We present an exact solution for a black hole localized near
an infrared wall in four-dimensional anti-deSitter space. By computing
the holographic stress tensor we show that the CFT dual of the black
hole is a 2+1-dimensional ball (i.e., a disk) of plasma at finite
temperature, surrounded by vacuum. This confirms some earlier
conjectures about plasma balls in AdS/CFT. We also estimate the value of
the surface tension for the ball. The solution displays a number of
peculiarities, most notably a non-trivial curvature of the boundary
geometry, as well as other properties associated to the vanishing
deconfinement temperature of the set up. We discuss how these features
are related to specific physics at the infrared and ultraviolet
boundaries for this solution, and should not be generic properties of
plasma balls. 

\end{abstract}

\end{titlepage}

\setcounter{footnote}{0}

%\tableofcontents

\newpage

\section{Introduction}

It has been shown recently that the Einstein equations that describe the
dynamics of black holes in anti-deSitter space can be recast, when the
length scales involved are much larger than the cosmological radius, in
the form of the equations of a relativistic conformal fluid
\cite{Bhattacharyya:2008jc}. This is likely to advance our understanding
of both of these apparently different physical systems. An interesting
extension of this correspondence involves fluids with surface
boundaries. Such is the case of plasmas in quantum field theories with a
confinement first order phase transition, where balls of deconfined
plasma surrounded by the confining vacuum can form 
slightly above the deconfinement temperature.

Gravitational duals for such plasma balls are expected in theories of
anti-deSitter gravity that admit solutions for homogeneous black branes
and solutions with an `infrared bottom', \ie a limit to the inward
extent away from the boundary in the Poincar\'e patch of AdS. Plasma
balls were discussed in detail in this context in \cite{Aharony:2005bm},
mostly within a purely gravitational realization of dual confinement,
but the phenomenon is expected to be present largely independently of
the details of the infrared cutoff. It was argued that the gravitational
dual of plama balls should be black holes in the bulk, localized near
the infrared end of spacetime. Their horizon should be `pancaked' and
well approximated by a black brane in the region away from the interface
with the infrared end. The correspondence has been subsequently
exploited in \cite{lumps} in order to reveal properties of hitherto
unknown black holes in AdS that exhibit behavior remarkably similar to
black holes in asymptotically flat spacetime.

This correspondence between plasma balls and black holes remains,
however, conjectural in some of its aspects. The gravity/fluid
correspondence has been proven in \cite{Bhattacharyya:2008jc} only for
plasma fluids without surface boundaries, since at the plasma boundary
the density and pressures vary rapidly and the hydrodynamic description
breaks down. Thus the dual bulk physics of the surface boundary requires
solving the full gravitational equations and the approximations in
\cite{Bhattacharyya:2008jc} are not valid. Ref.~\cite{Aharony:2005bm}
obtained a numerical solution for an infinite planar interface between
the AdS-soliton (confined) and black brane (deconfined) phases. But so
far no solution for a black hole dual to a \textit{finite} plasma ball
has been shown to exist\footnote{Ref.~\cite{Giddings:2002cd} obtained
some information on infrared black holes using linearized gravity around
an AdS infrared brane background.}. Our purpose is to present and
discuss an exact solution that describes such an object.

To this end, we use a previously known exact solution for an
accelerating black hole in four-dimensional AdS. Our construction can be
regarded as the complement of the one in \cite{Emparan:1999wa} for a
black hole localized in an ultraviolet brane. In
ref.~\cite{Emparan:1999wa}, the spacetime for a black hole moving in an
accelerated trajectory inside AdS was cut off in a Randall-Sundrum II
braneworld construction, throwing away the region from the brane to
infinity --- the ultraviolet\footnote{This and other related solutions
have subsequently been used as exact models for other configurations
relevant to the cutoff AdS/CFT correspondence
\cite{cads,Tanaka:2002rb,Emparan:2002px}.}. In the present paper we
will instead retain this region and cut out its complement --- the
infrared. This crude procedure provides the simplest way to implement an
infrared cutoff. We shall show that the black holes we construct give
rise to several of the properties conjectured for the dual plasma balls
--- actually plasma disks, since they live in 2+1 dimensions.

The solutions also display certain peculiarities related in one way or
another to the conditions imposed at the infrared bottom and at
asymptotic infinity. The way we introduce the infrared end by
effectively breaking spontaneously the conformal symmetry gives rise to
a number of subtleties, which affect the properties of the deconfinement
transition and possibly also the shape of the black hole
horizon. At the opposite end, we encounter issues that relate to the
asymptotic behavior of the solutions. The black hole
solutions in AdS$_4$ that we use were originally found by looking for
metrics in an algebraically special class (Petrov type-D) instead of
specifying any particular kind of asymptotics. Thus we find a somewhat
unconventional behavior of the geometry at the asymptotic AdS boundary,
which however may turn out to have an interesting interpretation. 

Our main objective in this paper is to establish the main features of
the black holes localized in the infrared and of their plasma ball
duals. We do this in sections \ref{sec:bulksoln} and \ref{sec:bdrydual}.
The subtleties of boundary conditions are discussed in section
\ref{sec:discuss}. However, since they are somewhat secondary to our
initial purpose here, they are not dealt with in full detail. Pursuing
these interesting matters further is left for future work.

\section{The exact solution}
\label{sec:bulksoln}

Our starting point is a subfamily of a class of solutions to Einstein's
equations with a negative cosmological constant in four dimensions first
presented in \cite{Plebanski:1976gy}. We discuss it succintly, since
more details can be
found in \cite{Emparan:1999wa}\footnote{The use of this solution to
construct black holes on an infrared brane was first discussed in
\cite{Emparan:1999wa}. Another reference using this idea is
\cite{Zakout:2002mg} but our analysis differs very substantially
from it.}. The metric is
\beq\label{soln}
ds^2=\frac{\ell^2}{(x-y)^2}\left[-\ell^{-2}
H(y)dt^2+\frac{dy^2}{H(y)}+\frac{dx^2}{G(x)}+G(x)d\phi^2\right]
\eeq	
with
\beq
H(y)=y^2(1+2\mu y)\,,\qquad G(x)=1-x^2-2\mu x^3\,.
\eeq
Here $\ell=\sqrt{-3/\Lambda}$ is the cosmological radius. The solution contains a
single dimensionless parameter $\mu$. We take it to lie in the range
\beq\label{murange}
0<\mu<\mu_c\equiv\frac{1}{3\sqrt{3}}\,,
\eeq 
so that $G(x)$ has three real roots $x_0$, $x_1$, $x_2$ satisfying
\beq\label{xorder}
-\frac{1}{2\mu}<x_0<x_1<0<x_2\,.
\eeq 
Other parameter ranges are either equivalent or do not give the physics
we seek. Although it is
customary in the literature to use $\mu$ as the parameter for this
family of solutions, we have found much more practical to use instead the root
$x_1$, in terms of which
\beq
\mu=\frac{1-x_1^2}{2x_1^3}
\eeq
and
\beq
x_0=x_1\frac{1+\sqrt{4x_1^2-3}}{2(x_1^2-1)}\,,\qquad 
x_2=x_1\frac{1-\sqrt{4x_1^2-3}}{2(x_1^2-1)}\,.
\eeq
The range \eqref{murange} and the ordering \eqref{xorder} are reproduced when 
\beq\label{x_1range}
x_{1c}\equiv-\sqrt{3}<x_1<-1\,.
\eeq
The upper limit $x_1\to -1$ corresponds to $\mu\to 0$, in which case the
coordinate transformation
\beq\label{xyzr}
x=-\frac{\ell-z}{\sqrt{r^2+(\ell-z)^2}}\,,\qquad
y=-\frac{\ell}{\sqrt{r^2+(\ell-z)^2}}
\eeq
brings the solution into the form of the Poincar\'e patch of empty AdS$_4$
\beq\label{pads}
ds^2=\frac{\ell^2}{z^2}\left(dz^2-dt^2+dr^2+r^2 d\phi^2\right)\,
\eeq
($-\ell/y$ and $\arccos(x)$ are polar coordinates in the plane $(r,z)$
centered at $(0,\ell)$).
The opposite, `critical' limit $x_1\to x_{1c}$ corresponds to the upper
bound on $\mu\to\mu_c$ in
which the black hole becomes, in a specific way that we explain below, of
infinite size.

If we demand that $G(x)>0$ so that $x$ and $\phi$ are spatial
coordinates, then we must require that $x$ lies in the range
$x_1\leq x \leq x_2$. We will shortly restrict this range further to
incorporate the infrared cutoff.

It is easily seen that there is an event horizon at $y=-1/(2\mu)$, and a
curvature singularity at $y=-\infty$. The range of $y$ outside the
horizon is $-1/(2\mu)<y<x$. Asymptotic infinity lies at $y=x$. The
temperature of the horizon, normalized
relative to the Killing vector $\partial_t$, is
\beq\label{Tbh}
T=\frac{1}{8\pi\mu\ell}\,.
\eeq

The locus $x=x_1$ is a semi-axis of rotation (a fixed-point set) for
$\partial_\phi$, which
extends from the horizon towards asymptotic infinity. In order to
avoid a conical singularity there we identify $\phi\sim\phi+\Delta\phi$,
with
\beq \label{Delphi}
\Delta\phi=\frac{4\pi}{G'(x_1)}=\frac{4\pi x_1}{x_1^2-3}\,. 
\eeq
Observe that $\Delta\phi>2\pi$ and that it diverges as $x_1\to -\sqrt{3}$.
The canonically normalized angular variable is
\beq
\tilde\phi=\frac{x_1^2-3}{2 x_1}\phi\,.
\eeq
with $\tilde\phi\sim \tilde\phi+2\pi$. This identification introduces a conical
excess-angle singularity along the opposite semi-axis $x=x_2$.
Physically, this line singularity is interpreted as a `strut' that pushes
the black hole and keeps it in a trajectory that accelerates towards the
boundary of AdS. We shall eliminate this singularity by removing the
portion of AdS where it lies. 

Indeed, we are interested in introducing an infrared cutoff in AdS
space. We shall implement this by cutting off the space at a surface
whose extrinsic curvature is proportional to its induced metric. Such a
cutoff can be regarded as due to a `wall' or `infrared brane' with a
distributional source of energy-momentum of vacuum type. 
These requirements for the infrared cutoff are met at the surface
$x=0$. So we cut off the geometry \eqref{soln} by restricting $x$ to the
range
\beq\label{xrange}
x_1\leq x\leq 0\,.
\eeq
In this way we retain the region of AdS from the wall out to infinity.
This is most clearly seen when the black hole is absent, $\mu=0$, where
the change \eqref{xyzr} maps the interval \eqref{xrange} of \eqref{soln}
to $0\leq z\leq\ell$ in \eqref{pads}. This is in contrast to the construction
in \cite{Emparan:1999wa}, where the complement of this region was kept.
In the following we shall often refer to this infrared cutoff as `the
brane' even if we do not intend to imply any particular connection to
braneworld models. 

The brane thus introduced has the gravitational effect of a
negative-tension domain wall. The unstable negative energy oscillations
can be projected out in a conventional manner by imposing `orbifold'
boundary conditions. On the other hand, the infrared cutoff introduces a
mass gap $\sim 1/\ell$ in the spectrum of Kaluza-Klein graviton
modes\footnote{We are using here terminology borrowed from braneworlds,
where a Kaluza-Klein-type mode decomposition of fields along the
direction $z$ is customarily performed.}. But there is no mechanism to
stabilize the location of the brane so there is a massless modulus, the
radion, corresponding to its position. In sec.~\ref{sec:discuss} we
discuss further the nature of this cutoff and the effects of the
massless radion.

\subsection{Horizon geometry}
\label{subsec:horgeom}

The spatial geometry induced on the horizon at $y=-1/2\mu$ is
\beq
ds^2_H=
\frac{\ell^2}{\left(x+1/(2\mu)\right)^2}\left[\frac{dx^2}{G(x)}+G(x)d\phi^2\right]\,,
\eeq
with $x$ within the range \eqref{xrange}. The coordinate $x$ plays the
role of (the cosine of) a polar angle along the horizon, with $x=x_1$
being
the pole, \ie the axis of rotational symmetry, and $x=0$ the point of
contact with the infrared brane. The horizon area is easily computed to
be
\beq\label{ahor}
A_H=4\pi\ell^2\frac{(x_1^2-1)^2}{x_1^2(3-x_1^2)}\,.
\eeq
This area grows monotonically as $\mu$ increases from $0$ to $\mu_c$, \ie $x_1$
decreases from $-1$ to $-\sqrt{3}$, and becomes much larger than
$\ell^2$ as $x_1$ approaches the critical value $-\sqrt{3}$. The latter
is the regime we will be
mostly interested in.

\begin{figure}%[th]
\centerline{\includegraphics{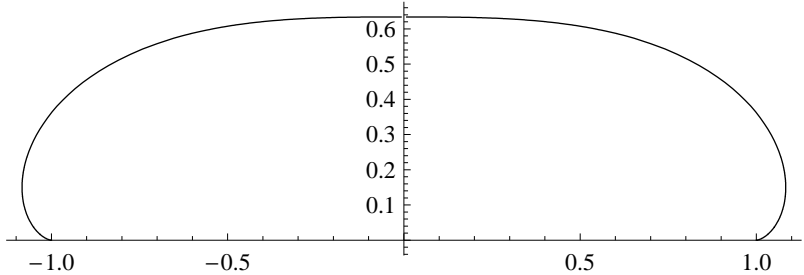}} \caption{\small 
Profile of the black hole localized near the infrared wall exhibiting
the `low-wetting' effect. Shown here is (a polar section of) the
embedding in Euclidean 3D space of the horizon geometry, cut off by the
infrared brane (which lies along the lower horizontal axis in the
figure). The figure corresponds to $\mu=1/(4\sqrt{2})$
($x_1=-\sqrt{2}$). As the black hole grows larger the horizon spreads
along directions parallel to the infrared brane, but it also bends over
near its edge before touching the brane (`droplet on a plastic sheet').
Units
correspond to $\ell=1$.} 
\label{fig:wet} 
\end{figure}

This horizon
displays what can be described as a \textit{low-wetting effect}. A way to
illustrate it pictorially is
by drawing the profile of an embedding of the horizon into 3D Euclidean
space, see fig.~\ref{fig:wet}. For very small $\mu$ the horizon is
approximately hemi-spherical, with radius $\simeq 2\mu\ell$ much smaller
than the AdS radius $\ell$. But as $\mu$ grows the horizon's
circumference is longest at a finite
distance away from the brane, like a droplet of water on a
plastic surface. 
The 3D embedding cannot be performed for
$x_1<-\sqrt{2}$ since in that case the intrinsic
scalar curvature of the horizon in a region near the rotation axis
becomes negative. Nevertheless, the low-wetting can be demonstrated by
showing that, for every given value of $x_1\in (-\sqrt{3},-1)$, the
circumference length of the horizon at latitude $x$
\beq
\mathcal{C}(x)=\Delta\phi\sqrt{g^{H}_{\phi\phi}(x)} 
\eeq 
reaches a maximum at a value $x_{m}$ away from the brane location $x=0$.
In the critical limit $x_1\to-\sqrt{3}$ the length $\mathcal{C}(x)$
diverges for any $x>x_1$, but the limiting $x$ for the maximum circle
approaches $x_m\simeq -0.54$ and the ratio between the maximum
circumference of the horizon and the circumference on the brane remains
finite\footnote{The exact values for $x_m$ and the ratio can be found
but are not too illuminating.} 
\beq
\frac{\mathcal{C}(x_{m})}{\mathcal{C}(0)}\simeq 1.107>1\,. 
\eeq 
This low-wetting effect is presumably related to the specific nature of
the infrared cutoff we have imposed and of its interaction with the
black hole horizon, which seems to be repulsive. This is in spite of the
fact that the negative-tension brane attracts objects in the bulk. The
latter can be regarded as the reason why the horizon flattens
along the brane, and is expected for any kind of infrared cutoff.
However, we do not find any reason why the low-wetting effect should be
generic for the horizons of duals of plasma balls.

A more striking feature is the appearance of a
region of negative curvature around the center of the horizon. A better
understanding of this important region is obtained from our next
analysis.

\subsection{Limit to a black brane}
\label{subsec:critlim}

Motivated by the dual physical picture of plasma configurations, we
expect that as the black hole becomes larger it should approach, in a
region around its center (rotation symmetry axis), the geometry of a
black brane, which is the dual of the deconfined phase. Something like
this does happen in our solution, but with an unexpected twist. 

In order to investigate the limit in which the black hole becomes large,
$\mu\to\mu_c$, we introduce
\beq\label{mueps}
\mu=\mu_c\left(1-\frac{\epsilon^2}{2}\right)
\eeq
for small $\epsilon$,
so that
\beq\label{xeps}
x_0=x_{1c}-\epsilon+O(\epsilon^2)\,,\qquad
x_1=x_{1c}+\epsilon+O(\epsilon^2)\,,\qquad
x_2=\frac{\sqrt{3}}{2}+\frac{\epsilon^2}{12\sqrt{3}}+O(\epsilon^3)\,.
\eeq
Furthermore, to focus on the geometry near the rotation axis of the
horizon $x=x_1$ and
away from the edges, we define a coordinate $\xi$,
\beq
x=x_{1c}+\epsilon\cosh\xi\,.
\eeq
that remains finite as $\epsilon\to 0$. It is also convenient to
introduce a new radial coordinate
\beq
\varrho=-\frac{\ell}{y+\sqrt{3}}\,.
\eeq
The horizon at $y=-3\sqrt{3}/2$ is now mapped to 
\beq
\varrho_H=\frac{2\ell}{\sqrt{3}}\,.
\eeq
In the limit $\epsilon\to 0$ the solution \eqref{soln} becomes a hyperbolic AdS black
hole (also known as `topological black hole') 
\beq\label{topobh}
ds^2\to -f(\varrho)dt^2+\frac{d\varrho^2}{f(\varrho)}
+\varrho^2\left(d\xi^2+\sinh^2\xi \,d\tilde\phi^2\right)\,,
\eeq
where
\beq
f(\varrho)=\frac{\varrho^2}{\ell^2}-1-\frac{2m}{\varrho}
\eeq
with mass parameter
\beq\label{topom}
m=\frac{\ell}{3\sqrt{3}}\,.
\eeq
This is different from the black brane (`planar black hole') that naively we might
have expected, which would have
$f=\frac{\varrho^2}{\ell^2}-\frac{2m}{\varrho}$ and a planar horizon.
Nevertheless, this geometry is also dual to a deconfined phase of the
CFT, with the deconfined degrees of freedom contributing a large
entropy (dual to the Bekenstein-Hawking entropy).

Hyperbolic black holes approach planar black holes when they are very
large, $m\gg \ell$. However we are not in this regime since in the
present case $m$ is
of the same order as $\ell$.\footnote{In fact the mass is precisely that for
which $f(\varrho)$ has a double zero at $\varrho=-\ell/\sqrt{3}$ (inside the event
horizon), $f(\varrho)=\frac{\ell}{\varrho}\left(\frac{\varrho}{\ell}+
\frac{1}{\sqrt{3}}\right)^2
\left(\frac{\varrho}{\ell}-\frac{2}{\sqrt{3}}\right)$.
While we do not have any physical explanation for this particular value,
one could anticipate it since the function $f(\varrho)$ is essentially
the same as $H(y)$, and $H(y)$ has a double zero at $y=0$ \ie
$\varrho=-\ell/\sqrt{3}$.} Observe also that the temperature remains
finite in this limit. Naively one would have expected it to approach the
deconfinement temperature, which, as we discuss in
sec.~\ref{sec:discuss}, is zero for this model. These are indications
that, although the horizon area diverges as $\mu\to\mu_c$, there is a
sense in which the black hole is not spreading indefinitely along the
infrared wall. The growing
lengths and areas in the geometry are a result of the coalescence of
two roots of $G(x)$, $x_0$ and $x_1$, which make the periodicity $\Delta\phi$ and
the proper distance along the polar direction, $dx/\sqrt{G(x)}$, grow
unbounded. But observe in particular that the latter diverges in the
region near the pole $x\approx x_1$, and not near the edge $x\approx 0$.
In this sense the black hole horizon does not extend to arbitrarily large
distances along the brane. Rather, its central area is warping to become
a hyperbolic horizon.

\bigskip

In summary, our black hole localized near the infrared wall demonstrates
some of the properties anticipated for such solutions on general grounds,
but also a number of peculiarities that were less expected. In the next
section we turn to the holographic dual description, which shows again
both kinds of aspects from a different perspective.

\section{The plasma ball}
\label{sec:bdrydual}

\subsection{Boundary Geometry and Holographic Stress Tensor}

The properties of the CFT state dual to our black hole
can be conveniently derived by writing
the metric in
Fefferman-Graham coordinates $(z, x^i$) in which
\begin{equation}\label{fg}
d s^2 = \frac{\ell^2}{z^2}\left(d z^2 + g_{ij}(z,x) d x^i d x^j\right)\,.
\end{equation}
When the metric
$g_{ij}$ is expanded in power series of $z$, 
\beq
g_{ij}(z,x)=\sum_{n=0}^\infty g_{ij}^{(n)}(x)\, z^n \,,
\eeq
the leading term $g_{ij}^{(0)}$ corresponds to (a representative of the
conformal class of) the metric at the boundary. Given $g_{ij}^{(0)}$ the
Einstein equations fix the terms $g_{ij}^{(1)}$ and $g_{ij}^{(2)}$
but $g_{ij}^{(3)}$ contains
information about the particular state of the solution at hand
\cite{Henningson:1998gx,de Haro:2000xn}. More specifically, it gives
the renormalized stress tensor of the dual theory as
\begin{equation}
\langle T_{ij}(x)\rangle = \frac{3\ell^2}{16\pi G} g^{(3)}_{ij}(x)\,.
\end{equation}

%Writing the exact metric \eqref{soln} in the form \eqref{fg} is not
%easy. We have seen that the change \eqref{xyzr} does the job when
%$x_1=-1$, but in general with $x_1\neq -1$ this change of coordinates
%introduces cross terms $g_{r z}$. To eliminate these we perform
%a further change of the $(z,r)$ coordinates, expanded in power series
%of $z$, in such a way that at each order in $z$ the term $g_{r
%z}^{(n)}$ vanishes. For our purposes it suffices to proceed only up to
%$n=3$. 
We defer the details of the calculation to the appendix, and quote here only
the final results. In order to reduce clutter we set $\ell=1$ in
eqs.~\eqref{bdryg}--\eqref{asympg0}.

The boundary metric
is
\beqa\label{bdryg}
g^{(0)}=-\frac{1+ x_1^2 \rho}{x_1^2(1+\rho)}dt^2+
\frac{(1+\rho)^4}{\rho(1+ x_1^2 \rho)(3-x_1^2+3\rho+\rho^2)}d\rho^2
+\frac{\rho(3-x_1^2+3\rho+\rho^2)}{x_1^2(1+\rho)}d\phi^2\,.
\eeqa
The coordinate $\rho$ runs from $0$ to $\infty$ and when $x_1=-1$ it is
related to $r$ in \eqref{pads} as $r=\sqrt{\rho(\rho+2)}$. Although
$\rho$ does not correspond to any invariant notion of radius it is a
convenient coordinate for keeping expressions as simple as possible.

The holographic stress tensor is
\beqa\label{bdryTij}
\langle T^t_t\rangle&=&\frac{1}{16\pi G}\frac{x_1^2-1}{(1+\rho)^3}
\left(
-3\frac{1+x_1^2\rho}{(1+\rho)^3}+1\right)\,,\nonumber\\
\langle T^\rho_\rho\rangle&=&
\frac{1}{16\pi G}\frac{x_1^2-1}{(1+\rho)^3}\,,
\\
\langle T^\phi_\phi\rangle&=&\frac{1}{16\pi G}\frac{x_1^2-1}{(1+\rho)^3}
\left(
3\frac{1+x_1^2\rho}{(1+\rho)^3}-2
\right)\,.\nonumber
\eeqa
This stress tensor must be traceless and conserved,
$g^{(0)}_{ij}\langle T^{ij}\rangle=0$ and $\nabla_i\langle
T^{ij}\rangle=0$ \cite{de Haro:2000xn}. We have explicitly checked that
these properties hold.

The first thing to observe is that the boundary metric is not flat. More
precisely, since the Cotton tensor of $g^{(0)}$ does not vanish the
conformal class of the boundary metric is not flat. Perhaps more
strikingly, we get a different boundary metric for each value of $x_1$.
In the usual AdS/CFT interpretation, this means that in the bulk there
are not only normalizable excitations of the graviton, which give rise
to $\langle T_{ij}\rangle$, but also non-normalizable gravitational
modes which deform the boundary geometry. Since our solutions only have
one parameter, $x_1$, we cannot change independently the normalizable
and non-normalizable contributions. Thus, the stress tensor $\langle
T_{ij}\rangle$ that we obtain for each bulk black hole corresponds to a
state of the dual CFT living in a different background spacetime. We
will comment on a possible alternative view of this effect in section
\ref{sec:discuss}.

The metric near the origin is
\beq\label{origing0}
g^{(0)}\to
-\frac{dt^2}{x_1^2}+\frac{1}{3-x_1^2}\frac{d\rho^2}{\rho}+\frac{3-
x_1^2}{x_1^2}\rho d\phi^2\qquad (\rho\approx 0)\,.
\eeq
The change
$\rho=\tilde\rho^2/4$ reveals more clearly that this is a flat
metric when the periodicity of $\phi$ is fixed to
\eqref{Delphi} in order to avoid a conical singularity at the origin.
This has the consequence that the metric near infinity
\beq\label{asympg0}
g^{(0)}\to
-dt^2+\frac{1}{x_1^2}\left(d\rho^2+\rho^2 d\phi^2\right)\qquad
(\rho\to \infty)\,,
\eeq
is only locally asymptotically flat: it asymptotes to a
cone with excess angle $\Delta\phi-2\pi>0$. Observe also that the origin
is redshifted by a factor $1/|x_1|<1$ relative to infinity.

In order to characterize the radial profile of the configuration we
shall use, as an appropriate invariant quantity, the proper
radial distance
\beq
r_p=\int_0^{\rho}d\rho'\sqrt{g^{(0)}_{\rho\rho}(\rho')}\,.
\eeq
Even if $g^{(0)}_{\rho\rho}$ diverges at $\rho=0$ in \eqref{origing0}, this is
simply a coordinate artifact and for $x_{1c}<x_1<-1$ the proper radius $r_p$
runs from 0 at $\rho=0$ out to infinity as $\rho\to\infty$. We could
also work with the circumferential radius $
\sqrt{g^{(0)}_{\phi\phi}(\rho)}\,\Delta\phi/2\pi$, but this contains a strong
distortion from the divergence of $\Delta\phi$ as $x_1\to-\sqrt{3}$.

\begin{figure}[t]
\centerline{\includegraphics[width=17cm]{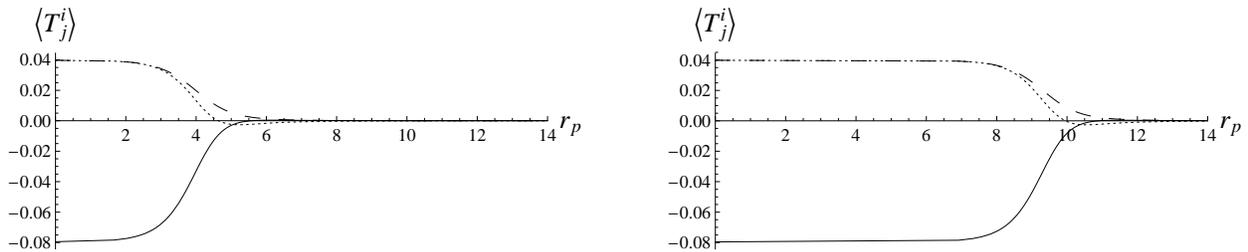}}
\caption{\small Holographic stress tensor components $\langle T^t_t
\rangle$ (solid), $\langle T^\rho_\rho\rangle$ (dashed), $\langle
T^\phi_\phi\rangle$ (dotted) of the plasma ball as a function of proper
radial distance $r_p$. Left: $x_1=x_{1c}+10^{-3}$. Right: $x_1=x_{1c}+10^{-7}$.
As $x_1$ approaches $x_{1c}$ 
the conformal thermal fluid relations $\langle T^t_t\rangle:\langle
T^\rho_\rho\rangle:\langle T^\phi_\phi\rangle=-2:1:1$ hold to great
accuracy in a larger region. The entropy of the dual black hole can be
reproduced by the entropy of a step-function homogenous plasma that
would extend out to $r_p=4.16$ (left), $r_p=9.48$ (right). 
Units correspond to $\ell=1$, $G=1$.}
\label{fig:Tijrp}
\end{figure}

To demonstrate the presence of the plasma ball we plot in fig.~\ref{fig:Tijrp}
the stress tensor as a function of proper
radius $r_p$. 
At the center of the ball we have
\beq
\langle T^i_j(r_p=0)\rangle =\frac{x_1^2-1}{16\pi G\ell}\mathrm{diag}(-2,1,1)\,,
\eeq
which is of the form of a perfect fluid of thermal conformal radiation.
Fig.~\ref{fig:Tijrp} shows
that as $x_1$ approaches the critical limit these central
values for the energy densities and pressures remain almost constant out
to a large radial distance, and then
drop fairly rapidly to zero. This is the profile expected for a ball of
plasma.

\begin{figure}[t]
\centerline{\includegraphics[width=16.5cm]{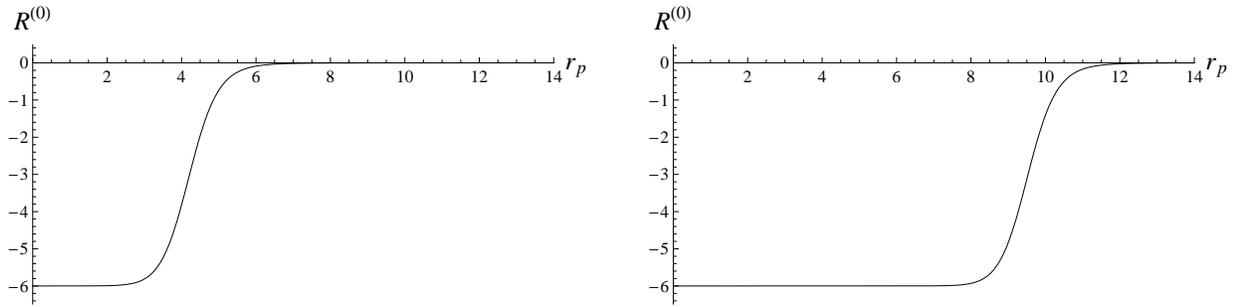}}
\caption{\small Scalar curvature $R^{(0)}$ of the boundary geometry as a
function of proper radial distance $r_p$. Left: $x_1=x_{1c}+10^{-3}$.
Right: $x_1=x_{1c}+10^{-7}$. The plasma of
fig.~\ref{fig:Tijrp} lies within the region of almost uniform scalar
curvature $R^{(0)}\simeq -6$. Units correspond to $\ell=1$.}
\label{fig:Rrp}
\end{figure}

There are, however, some peculiarities to this configuration of the dual
CFT that are the boundary counterpart of our previous result that as
horizons become large they develop a negatively curved region around
their rotation axis. The scalar curvature $R^{(0)}$ for the boundary
metric $g^{(0)}$ is negative. Fig.~\ref{fig:Rrp} shows that as $x_1$
approaches the critical limit there is a large region around the center
with almost uniform negative scalar curvature, which then turns fairly
rapidly into an almost flat region that extends to infinity. 

The region where the plasma resides coincides very approximately with
this region of negative scalar curvature $R^{(0)}\approx -6\ell^{-2}$ . This is
expected, given that the black brane limit of the bulk black hole, \ie
the dual of the deconfined phase, yields a hyperbolic black hole. Indeed
we can take the boundary counterpart of the limit in sec.~\ref{subsec:critlim} with
\eqref{xeps} and
\beq
\rho=\epsilon\frac{2}{\sqrt{3}}\sinh^2\frac{\xi}{2}
\eeq
to find that
\beq\label{limitg0}
g^{(0)}\to \frac{1}{3}\left(-dt^2+d\xi^2+\sinh^2\xi d\tilde\phi^2\right)
\eeq
and
\beq\label{limittij}
\langle T^i_j\rangle\to \frac{1}{8\pi G\ell}\mathrm{diag}\left(-2,1,1\right)\,.
\eeq
The stress tensor for the hyperbolic black hole of \eqref{topobh}, with
the mass given by \eqref{topom}, is $1/3^{3/2}$ of \eqref{limittij}
\cite{Emparan:1999gf}. The difference is accounted for by the conformal
factor $1/3$ in front of \eqref{limitg0}, which is due to a slightly
different choice of the Fefferman-Graham coordinate $z$ than would be
natural for the metric \eqref{topobh}. Taking this
into account, it follows that the temperature of the plasma ball is equal
to the Hawking temperature of the black hole in the bulk \eqref{Tbh}.

It is also curious to observe that outside the ball $\langle
T^t_t\rangle$ and $\langle T^\phi_\phi\rangle$ change sign, \ie there
are small negative energy densities and pressures, and in fact
asymptotically the ratios $\langle T^t_t\rangle:\langle
T^\rho_\rho\rangle:\langle T^\phi_\phi\rangle =1:1:-2$ are approached.

\subsection{Size and shape of the plasma ball}

We may now try to approximate the plasma ball by a
step-function distribution of homogeneous plasma  with
stress tensor equal to \eqref{limittij} extending out to a proper radial
distance $r_\mathrm{ball}$.
It is then also possible to assign an entropy to
this approximate plasma ball
as the local entropy
density of the homogeneous plasma times the area in $g^{(0)}$ occupied by the plasma.

The entropy density of the homogeneous plasma in \eqref{limittij} is computed from
the entropy density of the hyperbolic black
hole \eqref{topobh}, which is $\varrho_H^2/(4G\ell^2)=1/(3G)$, times a factor
$3=(\sqrt{3})^{2}$ that
takes into account the
conformal factor in \eqref{limitg0} relative to the boundary of
\eqref{topobh} as explained before. This is
\beq
s_\mathrm{plasma}=3\frac{1}{3G }=\frac{1}{G}\,.
\eeq
The area that the plasma occupies is
\beq
A_\mathrm{ball}=\ell^2\int_0^{\Delta\phi} d\phi\int_0^{\rho_\mathrm{ball}}
 d\rho\, \sqrt{g^{(0)}_{\rho\rho} g^{(0)}_{\phi\phi}}\simeq
\frac{4\pi \ell^2}{3-x_1^2}\rho_\mathrm{ball}\,,
\eeq
for $x_1$ close to $x_{1c}=-\sqrt{3}$. 
We may now fix the coordinate radius of the ball $\rho_\mathrm{ball}$ so
that the plasma entropy
\beq
S_\mathrm{ball}\simeq s_\mathrm{plasma}A_\mathrm{ball}\simeq
\frac{4\pi\ell^2}{G}\frac{\rho_\mathrm{ball}}{3-x_1^2}\,
\eeq
reproduces
the
Bekenstein-Hawking entropy of the bulk black hole. From \eqref{ahor} the latter is
\begin{equation}
S_{BH} =\frac{ A_H}{4G}\simeq \frac{4\pi\ell^2}{3G}\frac{1}{3-x_1^2}\,,
\end{equation}
again for $x_1\simeq x_{1c}$. Thus demanding that $S_\mathrm{ball}\approx S_{BH}$
fixes
\beq\label{rhoball}
\rho_\mathrm{ball}\approx \frac{1}{3}\,.
\eeq
The ($x_1$-dependent) value of $r_\mathrm{ball}$ that follows from this
does not admit a simple expression. However, if our definition of it is
any reasonable, $r_\mathrm{ball}$ should bound the region where
the profile of the
actual plasma is
almost constant.
Numerical inspection (see fig.~\ref{fig:Tijrp}) reveals that the proper
radial extent of the ball,
$r_\mathrm{ball}$, for the coordinate radius
\eqref{rhoball} coincides very
approximately with the point at which either of the components of $\langle
T^i_j\rangle$, as a function of $r_p$, has an inflection point
\beq\label{defrb}
\left. \frac{d^2\langle T^i_j\rangle}{d r_p^2}\right|_{r_p=r_\mathrm{ball}}\approx0\,.
\eeq
This shows that the approximation of the plasma ball as an
almost-homogeneous distribution of plasma is indeed sensible.

It is easy to see that the area occupied by the plasma ball, and
its proper radial extent out to $r_\mathrm{ball}$, grow unbounded near
the critical limit. However, in parallel with what we found for the bulk
horizon, this does not mean that the plasma ball extends out to
infinity. Since $\rho_\mathrm{ball}$ remains finite as $x_1\to x_{1c}$,
it follows that the divergence of the proper radius $r_\mathrm{ball}$ in
this limit is not due to an unbounded growth of the distance from any
point in the ball out to its edge. Rather, as eq.~\eqref{origing0} makes
clear, what happens is that the radial distance from any point {\it to
the origin}, 
\beq
r_p\simeq 2\ell\sqrt{\frac{\rho}{3-x_1^2}}\qquad (\rho \approx 0),
\eeq
diverges as $x_1\to -\sqrt{3}$. This is, the geometry around the origin
develops a trough within which the plasma ball lies, and this trough
gets deeper as $x_1$ approaches $x_{1c}$. 

So it is not quite accurate
to say that in this limit the ball spreads out to larger radii in the
surrounding vacuum. A more precise way to express this is through
the trace of the extrinsic curvature $K$ of the circle at the ball
radius\footnote{The extrinsic curvature is
computed for the circle at
$r_\mathrm{ball}$ in the entire spacetime, \ie not only in a spatial
section.}. In the limit $x_1\to x_{1c}$ this curvature at the
ball's outer circumference remains finite,
\beq\label{extK}
\left. K\right|_{r_p=r_\mathrm{ball}}\to\; \sim 1.5\ell^{-1}
\eeq
so the edge of the ball always remains a circle of finite curvature and
does not approach the shape of a straight domain-wall. 

Thus the region occupied by the plasma does
become arbitrarily large but only within the negatively-curved trough.
The transition region between vacuum and plasma
phases has
thickness of order $\ell$, which could be expected, but this is also
the scale of other characteristic lengths of the ball. In particular the
extrinsic curvature radius at $r_\mathrm{ball}$, and also the thermal
length of the plasma and the intrinsic curvature radius of the
background are always of the same order $\sim \ell$. So we never
are really in the hydrodynamic regime where variations of the plasma
occur on scales much longer than the thermal wavelength.

Usually the stress tensor of large plasma balls can be effectively split
into a volume term of deconfined plasma (as above) and a boundary term
with a surface tension $\sigma$. The fluid equations then relate, very
generally, the pressure drop across the surface to its extrinsic
curvature through the Young-Laplace
equation
\beq\label{YL}
P=\sigma K\,.
\eeq
Despite the fact that
we never reach the hydrodynamic regime where $K$ and the ball's
intrinsic curvature are much
smaller than the temperature, it
may still be useful to assign a surface tension to these balls through
eq.~\eqref{YL}. Taking $P$ to be the value at the center of the ball,
$P=1/8\pi G \ell$, and $K$ as in \eqref{extK}, we
obtain
\beq
\sigma\simeq 0.026 G^{-1}\simeq 0.07 \frac{\epsilon}{T}\,,
\eeq
where in the last expression $\epsilon$ and $T$ are the energy density
and temperature of the plasma. Given the uncertainties involved, these
numbers should not be trusted much more than as estimations of order of
magnitude.

\section{Discussion: Physics of the infrared and ultraviolet
boundaries} 
\label{sec:discuss}

Let us recapitulate the salient aspects of our study in
sec.~\ref{sec:bdrydual}. 
Fig.~\ref{fig:Tijrp} illustrates our claim that the duals of the
black hole solutions \eqref{soln} can be regarded as plasma balls.
However certain features demand further
explanation.
Near the critical limit we find large disks of
deconfined plasma. But these disks lie within a trough of negative
curvature in the 2+1 background geometry, whose size is different for
each plasma ball. The trough gets deeper as the critical limit is
approached, but it never extends out to larger distances in the
asymptotically flat part of the geometry. Instead the edge of the
plasma, with thickness of order $\ell$,
approaches a circle of finite curvature radius. Related to
this, the pressure difference inside and outside the ball does not
become arbitrarily small as the ball becomes larger, and the limiting
temperature does not seem to admit an interpretation as a deconfinement
temperature. In addition, even if deep down the trough the solutions
approach a homogeneous plasma configuration, its fluctuations may not
admit a proper hydrodynamic description since the thermal wavelength of
the plasma is of the same order as the curvature of the background.

In the following we make a number of observations concerning the
properties of the ultraviolet and infrared boundaries that we hope help
understand several of these issues.

\subsection{Asymptotic boundary behavior and dynamical gravity on the
boundary}

From the perspective of the gravitational bulk, it is worth recalling
that the metric \eqref{soln} was found without any special regard for
specific boundary conditions. Instead it arose from a general analysis
of Petrov type-D metrics of the Einstein-Maxwell theory with a
cosmological constant \cite{Plebanski:1976gy}. It is quite possible that
other solutions for black holes accelerating in AdS$_4$ exist with
different asymptotic behavior. Physically, this may be realized if the
black hole is accelerated by a different, even if more singular, pushing
rod, for which the linear energy density and pressure are different from
each other and possibly even non-uniform (unlike in a conical
singularity where they are equal and uniform). This should certainly
modify the asymptotic behavior and conceivably it will allow the
existence of solutions for infrared black holes with (conformally) flat
boundary metric. Sadly, such metrics cannot be of special algebraic type
D, and we do not know of any method that can possibly give them in exact
form\footnote{This is in spite of the presence of two commuting abelian
isometries in the expected solution. In four dimensions, inverse
scattering techniques allow to construct solutions of $R_{\mu\nu}=0$ for
black holes accelerated by more general (and more singular) rods than
conical defect lines, but with a cosmological constant the theory is not
known to be integrable.}. Thus, approximate analytic or numerical
methods seem to be the best way to investigate them. The latter applies
as well to the higher-dimensional cases.

{}From a viewpoint closer to the AdS/CFT interpretation, it would not be
inconsistent to think of the solution as the result of first fixing a
$\mu$-dependent boundary geometry $g^{(0)}$ and then solving the
gravitational bulk equations for a black hole localized near the
infrared cutoff. This would in fact be the conventional interpretation
in AdS/CFT in which the boundary metric is not a dynamical field. This
is in effect what we have been assuming in sec.~\ref{sec:bdrydual}. But
such a choice of boundary behavior certainly does not appear natural. 

Instead, it might appear more natural to think of $g^{(0)}$ as a
geometry modified by the presence of the plasma. This would require that
the boundary metric becomes a dynamical field. Interestingly, it has
been argued recently that dynamical gravity at the AdS boundary can
arise by appropriate choice of asymptotic boundary conditions
\cite{Compere:2008us,deHaro:2008gp}. Namely, the conventional Dirichlet
conditions on the graviton lead to non-dynamical gravity, as in standard
AdS/CFT, but mixed Dirichlet-Neumann conditions (with an appropriate
finite norm for the bulk graviton) can give rise to a suitable gravity
theory. In the case of AdS$_4$ the gravitational action is generically
expected to contain the Einstein-Hilbert term, a cosmological constant,
and a gravitational Chern-Simons term, each with an independent
coefficient, and a possible coupling to sources. 

The cosmological constant is absent in our solutions with asymptotically
(locally) flat $g^{(0)}$, but the other two terms should be present. The
Einstein-Hilbert term is a relevant operator and at large distances
should dominate the dynamics over the Chern-Simons term, which is
marginal and should kick in at shorter distances. Evidence for this
comes from the fact that, asymptotically as
$\rho\to\infty$, the Einstein equations
\beq\label{einsbdry}
R^{(0)}_{ij}-\frac{1}{2}g^{(0)}_{ij}R^{(0)}=\frac{8\pi G}{\ell}\langle T_{ij}\rangle
\eeq
are satisfied by \eqref{bdryg} and \eqref{bdryTij} to leading order in
$1/\rho$, \ie to linearized order around (locally) flat space. Observe
though that the linearized gravity dynamics implied in \eqref{einsbdry}
is not the leading effect at large distances: in \eqref{bdryg} there is
an asymptotic conical angle. The fact that it is an excess angle is
suggestive of a repulsive gravitational effect (possibly a negative
mass) and perhaps some kind of instability\footnote{Note the low-wetting
in sec.~\ref{subsec:horgeom} also suggests repulsion.
Refs.~\cite{Marolf:2001ne} discuss possibly related
instabilities of black holes on negative-tension branes}. On
the other hand, dynamical gravity on the boundary would make the
presence of a conical deficit angle at the origin (which we have barred,
but would appear if the asymptotic geometry were globally flat,
$\phi\sim\phi+2\pi$) more acceptable, since it could be caused by a
pointlike mass source.

The further investigation of the appropriateness of this interpretation
of the curvature of $g^{(0)}$
lies beyond the scope of the present paper, but we find it highly
suggestive.

Finally, dynamical gravity helps understand the puzzling issue of what
phase transition are these plasma balls associated to: at finite
temperature the vacuum would nucleate bubbles of plasma whose
self-gravity curves the spacetime that they appear in.

\subsection{Massless radion and the deconfinement transition}

We have introduced the infrared wall by simply cutting off the spacetime
at a surface whose extrinsic curvature is proportional to its induced
metric. This is a consistent procedure for creating a mass gap in the
graviton spectrum, but observe that since the introduction of this brane
does not modify the geometry away from it, it is obvious that it does
not affect the asymptotic behavior of the spacetime and does generate
any non-vanishing trace $\langle T^i_i\rangle$. Thus conformal symmetry
of the dual theory is not explicitly broken. Instead, it is
spontaneously broken. As discussed in sec.~\ref{sec:bulksoln}, the
position of the infrared wall in (empty) AdS is a modulus, commonly
known as the massless radion, which is the Goldstone boson of the
translation symmetry in $z$ that is spontaneously broken by the brane.
Hence the holographic dual theory contains a massless field whose vev
breaks spontaneously conformal invariance. 

The effects of this massless radion on the deconfinement phase
transition were investigated in \cite{Creminelli:2001th}. It was found
that at any small temperature the black brane is the dominant phase, \ie
the deconfinement temperature is vanishingly small (neglecting the
corrections of a thermally-generated potential for the radion). This is
not unexpected since the location of the infrared brane is
arbitrary and hence it cannot be said to lie above or below the horizon
of the black brane. Thus we recover essentially the same phase structure
as in the conformal theory.

Thus in our set up black branes form at any temperature above zero.
There is not a first-order phase transition, since the entropy jump to
the plasma phase can be made arbitrarily small (for finite volume) by
reducing the temperature. So proper plasma balls in equilibrum in flat
spacetime are not expected. Still, fireballs may form, say in
high-energy collisions, which would reproduce several of the aspects of
a black hole localized in the infrared end (see \cite{Giddings:2002cd}).
However, the ball would not have a surface tension to prevent it from
spreading in a flat spacetime and eventually cooling down to arbitrarily
small temperature.

The main mechanism that keeps our plasma balls in equilibrium seems to
be the curvature of the spacetime that creates a trough within which the
plasma lies. We have suggested above that this may be the result of
dynamical gravity on the boundary, and in this case the transition might
be of first order and thus a surface tension might perhaps be identified
as we have done. 

In order to introduce a proper confinement/deconfinement phase
transition (in a flat non-dynamical background) the location of the
infrared wall must be stabilized, \ie the radion must get a mass, and
the deconfinement scale will be set by this mass. Aspects of the phase
transition that results have been discussed in \cite{Creminelli:2001th}.

We do not see any easy way in which our solutions for infrared black
holes can be modified to accommodate for radion stabilization. Giving a
large mass to the radion makes the infrared brane very rigid and this
will presumably modify the shape and other properties of the black hole
that lies on top of it. In particular the low-wetting effect, which
apparently is due to the interaction between the black hole and the
infrared brane, could disappear when the dynamics of the latter is
modified. The stability properties of the black hole/infrared brane
configuration may also be altered.

\section{Outlook}

Finding new exact solutions for black holes on infrared branes with
different boundary behavior than in this paper, or in higher dimensions,
seems out of reach at present. Nevertheless, valuable information may be
obtained in a linearized approximation in which a specific asymptotic
boundary behavior is naturally imposed from the beginning. The study of
the linearized approximation to black holes in the infrared was
addressed in \cite{Giddings:2002cd}, which actually also considered the
effects of radion stabilization on the solutions. It seems possible to
us to extend this analysis to gain further information about black holes
that approach a brane-like behavior.

Other possible avenues for generalization of our solutions, using again
previously known exact metrics in AdS$_4$, involve rotation and charge
on the black holes. These are being currently investigated
\cite{sandro}.

\section{Acknowledgements} 

We are grateful to Shiraz Minwalla, Alex Pomarol, Mukund Rangamani and
Toby Wiseman for discussions, to Gary Horowitz and Rob Myers for
comments on a draft of this paper, and to Alessandro Maccarrone for
early collaboration. Part of this work was done while RE enjoyed the
warm hospitality of: Niels Bohr Institute; TIFR and ICTS during the 2008
Monsoon Workshop on String Theory in Mumbai; the CERN TH Institute
program on Black Holes; and the Cosmophysics group in KEK, Tsukuba. RE
was supported by DURSI 2005 SGR 00082, MEC FPA 2007-66665-C02 and CPAN
CSD2007-00042 Consolider-Ingenio 2010. GM was supported by MEC FPA
2007-66665-C01 and the ENRAGE Program MRTN-CT-2004-005616.

%%%%%%%%%%%%%%%%%%%%%%%%%%%%%%%%%%%%%%%%%%%
%\clearpage

\begin{appendix}

\section{Fefferman-Graham expansion of the metric}

In this appendix we set $\ell=1$.

In order to compute the holographic stress tensor we need to write the
metric \eqref{soln} in the form
\beq\label{fgexpz}
d s^2 = \frac{1}{z^2}\left(d z^2 + \sum_{n=0}^3 z^n\, g_{ij}^{(n)} 
d x^i d x^j +O(z^4)\right)\,. 
\eeq
For our purposes we need not care whether the $O(z^4)$ terms
include crossed $g_{iz}$, or $g_{zz}$ metric components.

We begin by introducing a `zeroth-order' set of coordinates
$(x,y)\to (r_0,z_0)$ through
\beq\label{xyzr0}
x=\frac{x_1(1-z_0)}{\sqrt{r_0^2+(1-z_0)^2}}\,,\qquad
y=\frac{x_1}{\sqrt{r_0^2+(1-z_0)^2}}\,
\eeq
(compare to \eqref{xyzr}). This brings the solution \eqref{soln} into
the form \eqref{fgexpz} only
when $x_1=-1$, as there appear non-zero $g_{r_0z_0}$ components at low
orders in $z_0$. In order to remove these up to order $z^3$ in an expansion
around $z=0$, we redefine the coordinates
\begin{eqnarray}\label{zrFGz0}
z_0 &=& z  +f_2 (\rho) z^2 + f_3 (\rho) z^3 + f_4 (\rho) z^4+O(z^5)\,,
\\\label{zrFGr0}
r_0 &=& \sqrt{\rho(\rho+2)}\left(1 + g_1(\rho) z 
+g_2 (\rho) z^2 + g_3 (\rho) z^3 + g_4 (\rho) z^4 +O(z^5)\right)\,.
\end{eqnarray}
The functions $f_n$, $g_n$ must vanish when $x_1=-1$ since this
corresponds to empty AdS. We have found convenient to introduce a
radial coordinate $\rho$ that looks awkward
for empty AdS where $r_0=\sqrt{\rho(\rho+2)}$ but which
simplifies significantly the results for generic $x_1$.

Plugging the coordinate transformations \eqref{xyzr0}, \eqref{zrFGz0} and
\eqref{zrFGr0} in the metric \eqref{soln} and expanding in powers of
$z$, at each order we require first that $g^{(n)}_{z\rho}=0$ and then
$g^{(n)}_{zz}=0$. In this way $g_n$ and $f_n$ are obtained as solutions
of purely algebraic equations, without the need of integrating
differential equations. We find
\begin{equation}
\begin{aligned}
f_2 =&-\frac{x_1^2-1}{2 (\rho +1)^3}\,,\\
f_3 =&\frac{x_1^2-1}{16 (\rho +1)^6}\left[3 (1-4 \rho ) x_1^2+8 (\rho
   +1)^3-15\right]\,,\\
   f_4=&-\frac{x_1^2-1 }{48 (\rho +1)^9}\Big[3 x_1^4 \left(20 \rho ^2-12
   \rho +1\right)-6 x_1^2 \left(10 \rho ^4+26 \rho ^3+18 \rho ^2-28
   \rho +3\right)+\\&+8 (\rho +1)^6-84 (\rho +1)^3+99\Big]
\end{aligned}
\end{equation}
and
\begin{equation}
\begin{aligned}
   g_1 =& \frac{1-x_1^2}{\rho ^2+3 \rho +2}\,,\\
   g_2=& \frac{x_1^2-1}{2
   (\rho +1)^4 (\rho +2)^2} \left[\rho ^4+5 \rho ^3-
    \rho ^2 \left(x_1^2-10\right)-\rho  \left(x_1^2-8\right)+x_1^2+1\right]\,,\\
   g_3=&\frac{\left(x_1^2-1\right)^2}{16 (\rho +1)^7
   (\rho +2)^3} \Big[8 \rho ^6+40 \rho ^5-8 \rho ^4
   \left(x_1^2-7\right)-4 \rho ^3 \left(3
   x_1^2+10\right)+\\&+\rho ^2 \left(29 x_1^2-169\right)+4 \rho 
   \left(11 x_1^2-29\right)-4 x_1^2+4\Big]\,,\\
   g_4=&-\frac{\left(x_1^2-1\right)}{96
   (\rho +1)^{10} (\rho +2)^4} \Big[12 \rho ^{10}
   \left(x_1^2-1\right)+8 \rho ^9 \left(13 x_1^2-10\right)-24
   \rho ^8 \left(3 x_1^4-18 x_1^2+2\right)+\\&+\rho ^7 \left(-390
   x_1^4+744 x_1^2+1392\right)+\rho ^6 \left(60
   x_1^6-465 x_1^4-1147 x_1^2+7030\right)+\\&+3 \rho ^5
   \left(42 x_1^6+569 x_1^4-2728 x_1^2+5615\right)-3
   \rho ^4 \left(131 x_1^6-2126 x_1^4+5326
   x_1^2-7489\right)+\\&+\rho ^3 \left(-1164 x_1^6+7329
   x_1^4-12824 x_1^2+15491\right)-6 \rho ^2 \left(85
   x_1^6-228 x_1^4+149 x_1^2-542\right)+\\&+12 \rho 
   \left(30 x_1^6-155 x_1^4+246 x_1^2-97\right)-4
   \left(3 x_1^6-27 x_1^4+41 x_1^2+7\right)\Big]\,.
\end{aligned}
\end{equation}
The results for $g^{(0)}_{ij}$ and $g^{(3)}_{ij}$ follow in a
straightforward way and are given in sec.~\ref{sec:bdrydual}. We have
checked that the other terms satisfy $g^{(1)}_{ij}=0$ and
$g^{(2)}_{ij}=R^{(0)}_{ij}-\frac{1}{4}R^{(0)}g^{(0)}_{ij}$ in agreement
with the general analysis in \cite{de Haro:2000xn}.

\end{appendix}

\end{document}